# Mobile Learning Environment System (MLES): The Case of Android-based Learning Application on Undergraduates' Learning

Hafizul Fahri Hanafi
*Computing Department Sultan Idris
Education University, Perak, Malaysia*
hafizul@fskik.upsi.edu.my

Khairulanuar Samsudin
*Computing Department Sultan Idris Education University,
Perak, Malaysia*
khairul@fskik.upsi.edu.my

*Abstract*—Of late, mobile technology has introduced new, novel environment that can be capitalized to further enrich the teaching and learning process in classrooms. Taking cognizance of this promising setting, a study was undertaken to investigate the impact of such an environment enabled by android platform on the learning process among undergraduates of Sultan Idris Education University, Malaysia; in particular, this paper discusses critical aspects of the design and implementation of the android learning system. Data were collected through a survey involving 56 respondents, and these data were analyzed by using SPSS 12.0. Findings showed that the respondents were very receptive to the interactivity, accessibility, and convenience of the system, but they were quite frustrated with the occasional interruptions due to internet connectivity problems. Overall, the mobile learning system can be utilized as an inexpensive but potent learning tool that complements undergraduates' learning process.

*Keywords* —mobile learning, android learning, teaching and learning

I. INTRODUCTION

Mobile technology has entered into the mainstream society, affecting the lives of many in recent years. This novel technology is slowly making its presence in the educational realm, which accords many opportunities to the learning and training. Its emergence in the educational world seems timely given the nature of today's learning requirements: wider, fast access to learning materials and persistent needs for prompt communication. Thus, learning institutions must seek every avenue for improvements to cope with new demands of teaching and learning process. Not surprisingly, new and emerging technologies are being passionately sought after by many institutions to provide better learning environments to their stakeholders, namely students and educators. One fine example of the adoption of new technologies is e-learning systems that have radically transformed learning–from being confined within the school walls to borderless landscape, empowering many trainees, students, pupils and other to learn with more academic rigor. With these formidable learning systems, more and more people can now seek informal education, irrespective of the academic background. Another benefit of such tools is that learning cost incurred unto students is drastically reduced as independent, self-paced learning can be done outside the schools and campuses. Based on these backdrops, it is eagerly anticipated that a new learning approach called mobile learning will be the next major enabler in this decade [1] that will take learning to another level as learning can be literally be conducted on our palms thorough wireless technologies [2]. Given the mass technological consumption of this new technology, a new learning paradigm will dawn over the academic horizon, bringing in new learning opportunities to all.

II. BACKGROUND AND RELATED WORKS

Mobile Learning or M-Learning is a type of e-learning that delivers educational contents and learning support materials through wireless communication devices [3]. Likewise, Traxler [4] describes mobile learning as a personalized, connected, and interactive use of handheld computers in classrooms, in collaborative learning during fieldwork, and in counseling and guidance. All these new learning activities are now possible through M-Learning which is empowered by recent advancements in mobile technology operating systems, notably the ubiquitous android platform. Android technology enables users to communicate with anyone at any time and place almost instantaneously transcending many barriers. As expected, mobile phones based on android platform have become an indispensable communication device for many people, particularly in younger segments of the population, such as school students. Android is an open source mobile operating system that has been supported by Google Corporation, the world leading search Engine Company. One major reason for the pervasive adoption of android in the mobile market is that mobile applications developed through android development technology is more efficient and effective compared to the other technologies, such as mobile Window or Symbian operating systems, producing fast, user friendly and appealing applications. As application system files running on android are freely distributed in its Application Market, which is easily accessed over the internet, more and more people are attracted to use this operating system for their mobile devices. Moreover, android-based applications can be run on virtually any personal computers through the android emulator; and this capability promotes the growth of android market globally, leaving behind many rivals in its trail.

### III. LEARNING FRAMEWORK

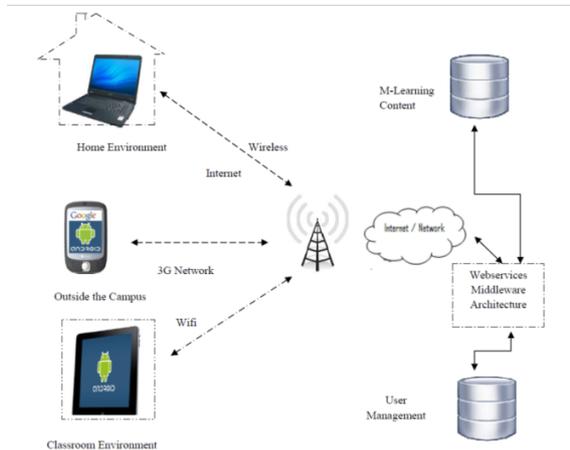

Figure 1 Mobile learning Application Learning Environment [3]

Mobile learning is developed by using multi platforms, languages, and technologies. Thus, learning can be carried out anywhere, anytime for as long as an institution's networking system can gain access to the wireless coverage [5] as illustrated in Figure 1. In this regard, android technology can help realize a mobile learning environment based on the network architecture shown with students gaining fast access to learning contents and materials of their studies by using their mobile phones.

This approach of learning is highly receptive to students as they are more likely to seek and use learning contents via mobile services rather than to find proprietary courseware that is not easily accessed. Propelled by the growing market of smart phones, M-learning is becoming more acceptable in teaching and learning process as these mobile devices are smart as they claimed to be – customizing their contents according to users' specific needs [6]. Teaching and learning has become more manageable and diverse as students can perform many learning activities freely and easily, for instance, they can download lectures notes almost instantaneously for lectures that they had missed. Predictably, mobile learning systems based on android technology are poised to dominate the M-learning realm given the rich, appealing multimedia contents such as audio, videos, animations that can be downloaded effortlessly into students' mobile devices.

### IV. LEARNING ENVIRONMENTS

Mobile learning is a form of digital learning which can be applied for teaching and learning purposes where some educational experts view it as a subset of e-learning but with a subtle difference—contents are delivered onto mobile devices rather than the ubiquitous desktop personal computers. Teaching and learning by using android platform can be easily implemented without heavy computing investment. There are several factors that make mobile computing as an appealing platform. First, android operating system to run the mobile devices is conveniently and freely available, thus making installation a simple, neat process.

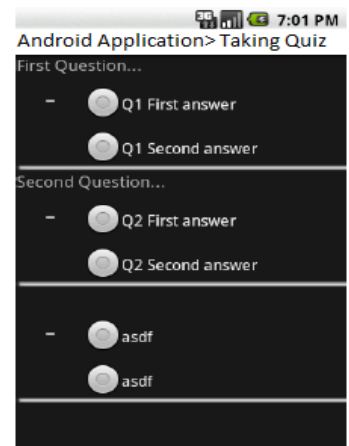

Figure 2 Example of Interface for Quiz

Second, there is a huge application base of learning materials and contents, which is continually expanding, that can be easily accessed by students and instructors alike. For example, students can download and practice short tests or quizzes on their mobile phones where prompt feedback is instantly displayed to improve comprehension This type of learning occurring in short bursts is appealing to young generation [7]. Figure 2 depicts such a test available on students' mobile phones that asks simple questions pertaining to a particular information technology course. In addition, students can download notes from Google doc website using android platform. Currently, the technology enables students to share and edit documents online collaboratively; thus, the notion of collective intelligence has transformed from an abstract concept into tangible realization in the educational realm. Figure 3 shown below illustrates the interface for students to download documents over the related website.

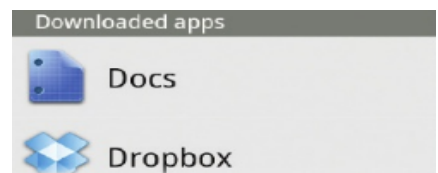

Figure 3 Google Doc

Third, there exists a repository of knowledge for sharing information among practitioners of mobile learning that contributes to the expansion of M-learning in campuses. This creates a community of practice that continually enhances the technical capabilities of M-learning systems irrespective of their background.

In spite of the many benefits accorded by M-learning, there are still some issues that need to be addressed by many concerned. For example, it is not that easy to measure the actual level of meaningful learning that takes place using this type of learning at best, and such a mode of learning can be easily abused at worst. Assimilating this technology into normal classroom activities requires not only structural changes but new thinking regarding learning of this nature is urgently required [8]. This problem is further compounded as individuals involved in M-learning come with different knowledge, skills, and expertise; and understandably, their involvements in this type of learning will be quite unpredictable to judge [9].

## V. RESEARCH METHODOLOGY

This study was conducted by means of a survey to measure undergraduates' perception of M-learning used throughout a semester long, beginning May 2011 until September 2011. Acknowledgement about the course contents and courses program was carried out by text messaging or SMS. The survey was conducted involving a total of 56 students were voluntarily participated in this study. They were divided into two different groups: the first group comprised undergraduates who used the university' e-learning; and the second group consisting of undergraduates who used their mobile phones, running on android 2.3.2, android 2.3.3 and android 2.3.4, to gain access to learning contents.

For the survey, the questionnaires were prepared in Google doc form, and the elicited data based on respondents' responses were collected from Google doc online database. For the learning materials, the two groups of students had to download notes from two different platforms. The first group downloaded learning materials from the e-learning system of the university, whilst the second group of students downloaded similar materials from Google Doc by using their mobile phones.

### A. Instrument of Research

The questionnaires comprised 4 sections pertaining to respondents' demographic, android learning environment system, e-Learning environment provided by the university, and the effectiveness of the android learning system. For the respondents' demographic, the data collected were gender, age, and academic achievement. For the Android learning environment, data gathered were related to respondents' perceptions on its features and usability. Likewise, the same perceptions of the respondents were elicited for the e-Learning environment system.

Respondents were asked for their opinions on the items of the questionnaire based on Likert-type scales as follows: 1 for strong disagreement, 2 for disagreement, 3 for being neutral, 4 for agreement, and 5 for strong agreement. All data collected were analyzed by using Statistical Package for Social Science (SPSS) version 12. Descriptive statistics for the demographic were based on frequency; for items pertaining to systems' features, mean scores were calculated to measure the respondents' responses.

### B. Findings and Discussion

Table I as shown below summarizes the descriptive statistics for the respondents demographic based on frequency counts.

TABLE I. DEMOGRAPHY OF THE STUDENTS INCLUDING THE ANDROID PLATFORM

| Item | Range | Frequency |
|---|---|---|
| Gender | Male | 20 |
|  | Female | 36 |
| Age (in years) | 18-22 | 10 |
|  | 22-26 | 46 |
| Respondents' academic achievements | CGPA < 3 | 15 |
|  | CGPA > 3 | 41 |
| Versions of Android Operating Systems | Gingerbread 2.3.2 | 10 |
|  | Gingerbread 2.3.3 | 20 |
|  | Gingerbread 2.3.4 | 26 |

Table II summarizes all the respondents' responses pertaining to their perceptions on the features provided by the novel learning systems, namely the android mobile learning system and the university's e-learning learning system.

TABLE II. RESPONDENTS' PERCEPTIONS ON THE LEARNING SYSTEMS' FEATURES

| Statements | Mean |
|---|---|
| 1. The university's e-learning environments is interesting and fun to use | 4 |
| 2. Android's mobile learning environment is interesting and fun to use | 4.5 |
| 3. Android's mobile learning environment provides more space for self-learning | 4.5 |
| 4. I can easily download lecture notes from Google Doc by using my mobile phone at any time and place | 5 |
| 5. I can easily download lecture notes from the university' e-learning system at any time and place | 3.5 |

All the five items of the questionnaires were analyzed to reveal mean scores based on the 56 respondents' responses. For items 1 and 2 (see Table II), the mean scores for the university's e-learning system and android mobile learning system were 4.0 and 4.5, respectively. Clearly, the respondents that had used the mobile learning system were more receptive to using the system where they regarded the system to be easy to use and also to be interesting. More poignant, all respondents that used the mobile platform were unanimous that they could use their hand phones for self-learning where item 3 of the questionnaire recorded a mean score of 4.5. This finding is not surprising as android-based mobile phones are quite affordable to own, and in terms of performance, they are very stable and could perform all the necessary functions with greater ease. These two factors seem to be a driving force to spur greater growth of mobile learning in the future.

For items 4 and 5 (see Table II), in terms of having the capability to download lecture materials, the mobile learning group and the e-learning group recorded mean scores of 5 and 3.5, respectively. Evidently, this feature of greater capability for downloading is better accorded by mobile learning platform compared to e-learning platform. Apparently, there are several reasons why mobile learning is greatly favored for this feature of a digital learning system. First, a learning system that guarantees uninterrupted access to learning materials can ensure smooth flow of learning process where students can download any documents at anytime, no matter where they are. This partly contributes to a more conducive learning environment that suits the needs of today's younger generation: the digital natives.

On the other hand, the availability of lecture materials by many e-learning systems is sometimes compromised by technical problems with most cases resulting in breakdowns, which hinder constant access for online documents. Thus, the feature of having full access of learning materials at all time will be a decisive factor that favors a mobile learning system over an e-learning system when the target groups of learning are those of adolescent age.

Another factor that seemed to motivate students to use the mobile learning system as compared to the university's e-learning system rested on the fact that the learning environment based on android itself was more interactive and simple to interact with. These students had ample time to download lecture notes from the Google Doc website without interruptions, and they could take quizzes and short tests at leisure, which further enriches their learning experience. Thus, learning becomes more fun. This is not unexpected as there is a vast repository of learning tools, widgets, and applications that could be accessed freely or bought at minimal cost from the android market. Once downloaded, all these digital materials can be utilized instantly and repetitively. Moreover, those students that have these materials can share with their peers by exchanging files through the Bluetooth technology, which is one of the standard features of today's mobile devices.

In sum, this research suggests that mobile learning can be quite easily implemented as clearly demonstrated by the respondents involved in this study. Mobile learning can be cost-effectively implemented as android operating system used to run the mobile phones is freely available. Moreover, newer, better versions of this operating system are constantly update, giving better performance in terms of processing and intuitive interface design. Mobile learning systems powered by android technology can make learning more fun, interactive and intuitive. This mobile learning system can be used by educational practitioners, such as instructors and teachers, to prepare the study notes in any standard digital formats and then upload them onto the Google Doc website, where they can be accessed and shared. Self-paced learning and collaborative learning can be realized with ease to improve the learning process that befits today's challenging learning environments.

## VI. FUTURE RESEARCH

Future research is needed to examine the full impact of mobile learning both from the technological and pedagogical perspectives. Expectedly, the introduction of any new, novel technology would have profound impact, affecting both students and educators. Students will be overwhelmed with the technological gizmo that is normally designed for routine chores, not for educational purposes. Thus, proper working ethics and code of practices are entailed to ensure optimal use of mobile devices for mobile learning. Likewise, educators must keep abreast with latest technologies to make efficient use of them. Therefore, future research should focus some of these issues to help realize digital learning environments that complement the conventional learning approach.

## VII. CONCLUSION

The use of mobile technologies, in particular the non-proprietary android technology, offers many educational opportunities to the stake holders: the students, the instructors, and the administrators. However, as for today, there are many emerging information and communication technologies entering the educational realm that forces practitioners to rethink how this novelty can be judiciously applied to improve the overall learning process. Many educational benefits of the novelty can be easily identified; however, realizing these is not a straightforward process as there is a web of interrelated factors that needs delicate unweaving to ensure effective and efficient implementations in educational institutions.

REFERENCES

[1] Keegan, D. (2002). The future of learning: From e-learning to m-learning.
 Retrieved September 7th, 2002 from the World Wide Web: http://learning.ericsson.net/leonardo/thebook/chapter4.html#milearn
[2] Barbosa and Geyer (2005). Pervasive personal pedagogical agent: A mobile agent shall always be a learner. Proceeding IADIS International Conference Mobile Learning, Malta 281-285 Bobnano
[3] Brown, H.T(2005),"Towards a model for MLearning", *International Journal on E-Learning*, 4(3),299-315
[4] Traxler, J. (2005). Institutional issues: Embedding and supporting. In A. Kukulska-Hulme & J.Traxler (Eds.), Mobile learning: A handbook for educators and trainers (pp. 173-188), London:Routledge.
[5] Shanmugapriya M.& Tamilarasia, A.(2011), Designing an m-learning application for ubiquitous learning environment in the android based mobile devices using web services, Indian Journal of Computer Science and Engineering(IJCSE) ,22-30
[6] Williams, A.J &Pence, H.E (2011). Smart phones, a powerful tool in Chemistry classroom, Journal of Chemical Education, 88, 683-686.
[7] P. Pocatilu, F. Alecu and M. Vetrici, Measuring the Efficiency of Cloud Computing for E-learning Systems, WSEAS TRANSACTIONS on COMPUTERS, Issue 1, Volume 9, January 2010, pp. 42-51.
[8] Winters, N.& Mor.Y (2008)'IDR: a participatory methodology for interdisciplinary design in technology enhanced learning' (Computers & Education, 50(2),579-600))
[9] F. Alecu, P. Pocatilu and S. Capisizu, WiMAX Security Issues in E-Learning Systems, Proc. of 2nd International Conference on Security for IT & C in Journal of Information Technology and Communication Security, Bucharest, November 2009, pp. 45-52

## AUTHORS PROFILE

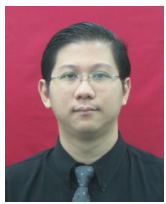

**Hafizul Fahri Hanafi** is a lecturer in the Computer Department of Sultan Idris Education University, Malaysia. He specializes in Software Engineering and E-Learning Technologies
(Email: apiltzs@gmail.com)

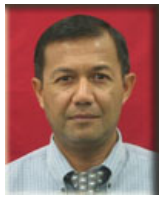

**Khairulanuar Samsuddin** is a lecturer in the Computer Department of Sultan Idris Education University, Malaysia. He specializes in Virtual Reality in Education and Engineering
(Email: khairul@fskik.upsi.edu.my)